\documentclass[manuscript]{aastex}

\newcommand{\Hipp}{{\it Hipparcos}}        

\newcommand{\HST}{{\it HST}}

\newcommand{\ltabout}{\, {}^<_\sim \,}

\def\pmb#1{\setbox0=\hbox{#1}
  \kern-.02em\copy0\kern-\wd0
  \kern.01em\copy0\kern-\wd0
  \kern.01em\copy0\kern-\wd0
  \kern.01em\copy0\kern-\wd0
  \kern.01em\copy0\kern-\wd0
  \kern-.02em\raise.01em\box0 }

\def\ref#1#2{$^{#1}$}

\shorttitle{Parallax of SS Cyg}
\shortauthors{Nelan \& Bond}

\begin{document}

\title{On the Hubble Space Telescope Trigonometric Parallax of the Dwarf Nova
SS~Cygni\altaffilmark{1}}

\author{Edmund P. Nelan\altaffilmark{2}
\and
Howard E. Bond\altaffilmark{2,3}
}

\altaffiltext{1} {Based in part on observations made with the NASA/ESA {\it
Hubble Space Telescope}, obtained from the data archive at the Space Telescope
Science Institute. STScI is operated by the Association of Universities for
Research in Astronomy, Inc., under NASA contract NAS~5-26555.}

\altaffiltext{2}
{Space Telescope Science Institute, 
3700 San Martin Dr.,
Baltimore, MD 21218, USA;
nelan@stsci.edu}

\altaffiltext{3}
{Department of Astronomy \& Astrophysics, Pennsylvania State University,
University Park, PA 16802, USA; heb11@psu.edu}

\begin{abstract}

SS Cygni is one of the brightest dwarf novae (DNe), and one of the best-studied
prototypes of the cataclysmic variables. Astrometric observations with the Fine
Guidance Sensors (FGS) on the {\it Hubble Space Telescope\/} (\HST\/), published
in 2004, gave an absolute trigonometric parallax of $6.06\pm0.44$~mas. However,
recent very-long-baseline interferometry (VLBI), obtained during radio outbursts
of SS~Cyg, has yielded a significantly larger absolute parallax of
$8.80\pm0.12$~mas, as well as a large difference in the direction of the proper
motion compared to the \HST\/ result. The VLBI distance reduces the implied
luminosity of SS~Cyg by about a factor of two, giving good agreement with
predictions based on accretion-disk theory in order to explain the observed DN
outburst behavior. This discrepancy raises the possibility of significant
systematic errors in FGS parallaxes and proper motions. We have reanalyzed the
archival \HST/FGS data, including (1)~a critical redetermination of the
parallaxes of the background astrometric reference stars, (2)~updated input
values of the reference-star proper motions, and (3)~correction of the position
measurements for color-dependent shifts. Our new analysis yields a proper motion
of SS~Cyg that agrees well with the VLBI motion, and an absolute parallax of
$8.30\pm0.41$~mas, also statistically concordant with the VLBI result at the
$\sim$$1.2\,\sigma$ level. Our results suggest that \HST/FGS parallaxes are free
of large systematic errors, when the data are reduced using high-quality input
values for the astrometry of the reference stars, and when instrumental
signatures are properly removed.

\end{abstract}



\section{Introduction: HST Measurements of Trigonometric Parallaxes}

The Fine Guidance Sensors (FGSs) on the {\it Hubble Space Telescope\/} (\HST\/)
are a set of three interferometers used for guiding control of the telescope
during imaging or spectroscopic observations. In addition, the FGS system itself
can provide high-precision astrometry by measuring positions of a target star
and several surrounding astrometric reference stars with one FGS while the other
two guide the telescope. This capability of determining accurate trigonometric
parallaxes, with precisions that can be better than $\pm$0.2~mas, has been
applied to a variety of astrophysical problems, such as the zero-point of the
Cepheid period-luminosity relation (Benedict et al.\ 2007) and the ages of the
oldest stars in the solar neighborhood (Bond et al.\ 2013).

Recently, however, the possibility of significant systematic errors in \HST\/
trigonometric parallaxes has been raised by long-baseline interferometric radio
observations of the archetypical dwarf nova (DN) SS~Cygni, which gave very
different parallax and proper-motion (PM) results from the FGS measurement. DNe
are a subclass of cataclysmic variables (CVs), which are very close binaries
containing a low-mass main-sequence star that transfers material to an accretion
disk surrounding a companion white dwarf. In a DN system, at semi-regular
intervals, the accretion disk becomes unstable and brightens dramatically,
producing a DN outburst. 

\section{The Discrepant Radio and HST Parallaxes of SS~Cygni}

SS~Cyg at maximum light is one of the brightest DNe, and it is one of the
best-studied, having nearly continuous coverage of its optical variability since
its discovery in 1896 by Louisa Wells (as reported by Pickering 1896). These
brightness estimates---almost all of them by amateur astronomers---show that the
eruptions recur about every 49~days. During its outbursts, SS~Cyg becomes a
detectable radio source. This property has been exploited by Miller-Jones et
al.\ (2013, hereafter MJ13), who used very-long-baseline interferometry (VLBI)
to obtain astrometric measurements of SS~Cyg during five eruptions over the
years 2010 to 2012. The VLBI observations, triggered by notifications of
outbursts from amateur astronomers of the American Association of Variable Star
Observers (AAVSO), were made with the Very Long Baseline Array and the European
VLBI Network. The resulting radio-based trigonometric parallax, which is
referred to extragalactic calibrators, is an absolute parallax. By contrast, FGS
parallaxes from \HST\/ must be referred to nearby background stars, making it
necessary to estimate the distances to the reference stars in order to convert
the relative parallax of the target to absolute.

Harrison et al.\ (1999, hereafter H99) used \HST\/ to measure the trigonometric
parallax of SS~Cyg (and two other DNe), based on FGS data obtained in 1997 and
1998. They slightly revised their analysis of the data in a subsequent paper
(Harrison et al.\ 2000, hereafter H00), and again four years later (Harrison et
al.\ 2004, hereafter H04).  H04 give an SS~Cyg parallax of $6.06\pm0.44$~mas,
corresponding to a nominal distance of $165\pm12$~pc.

The VLBI parallax of SS~Cyg reported by MJ13, which is $8.80\pm0.12$~mas
($d\simeq114\pm2$~pc), is significantly larger than the \HST\/ value reported by
H04. MJ13 discuss several possible reasons for this large discrepancy, including
very large errors in one or more of the estimated parallaxes of the \HST\/
background reference stars, and statistical biases in the \HST\/ parallax
measurement. MJ13 point out that their 114~pc distance is in excellent agreement
with a distance predicted on the basis of models of unstable accretion disks,
whereas at the larger distance found by H04, with the implied doubling of the
system luminosity, the disk models predict that SS~Cyg would be a nova-like CV
in a state of nearly continuous outburst, rather than a DN as observed (see,
e.g., Schreiber \& G\"ansicke 2002 and Schreiber \& Lasota 2007).

\section{Reanalysis of the HST FGS Astrometry of SS~Cygni}

In order to investigate this discrepancy, we have made a new analysis of the
\HST\/ FGS observations of SS~Cyg.  The data were originally obtained under
\HST\/ programs GO-6538 and -7492, with observations being made near dates of
maximum parallax factors. Two spacecraft orbits each, separated by about 7~days,
were used in 1997 June, 1997 December, and 1998 December. In addition to SS~Cyg
itself, five nearby reference stars were observed several times during each
orbit, using FGS3 in POSITION mode. Further details of the observations are
given in col.~1 of Table~1.

We retrieved the FGS data from the \HST\/ archive\footnote{The \HST\/ data
archive is available at http://archive.stsci.edu/hst} and processed them through
the standard FGS calibration pipeline. The pipeline extracts the astrometric and
photometric measurements, computes the positions of the stars in the FGS
coordinate frame, and corrects the positions for differential velocity
aberration, optical field angle distortion (McArthur et al.\ 2002), and
spacecraft jitter and drift.  

Our analysis procedure is essentially the same as described in Bond et al.\
(2013). The data from the six \HST\/ orbits were combined, using a
four-parameter overlapping-plate technique that simultaneously solves for scale,
translation, rotation, and the PM and parallax of each star. These reductions
are  carried out using the least-squares program GAUSSFIT (Jefferys et al.\
1988).  We input the estimated reference-star parallaxes and PMs as observations
with errors, allowing the GAUSSFIT model flexibility to adjust their values
within the errors to minimize the solution's global errors. The output parallax
for SS~Cyg is thus absolute. No priors were assumed for SS~Cyg itself. Our
analysis differed from that of H04 in several additional respects, as follows.

\subsection{Input Estimated Parallaxes of the Reference Stars}

The five reference stars lie within a few arcminutes of SS~Cyg. We use the
designations that were assigned in the original proposal; the formatted \HST\/
proposal listings\footnote{Available at
http://www.stsci.edu/hst/phase2-public/7492.pro} may be consulted for the
coordinates and other details of the reference stars and observations.

H00 gave ground-based $VRI$ photometry for the reference stars. However, there
is additional photometry available in the literature for these stars, from the
following sources: (1)~AAVSO charts and associated magnitude
sequences\footnote{Available at http://www.aavso.org/variable-star-charts}
(2~stars); (2)~the AAVSO Photometric All-Sky Survey\footnote{Available at
http://www.aavso.org/apass} (APASS; Henden et al.\ 2012; all 5~stars); (3)~Grant
\& Abt (1959; 1~star); (4)~Henden \& Honeycutt (1997; 3 stars); (5)~Misselt
(1996; 2 stars); and (6)~the Tycho catalog (H{\o}g et al.\ 2000; 1~star). We
averaged the $B$, $V$, and $I$ magnitudes from these sources, giving equal
weight to each publication. Table~2 lists the star designations, the $V$
magnitudes, and the $B-V$ and $V-I$ colors, in cols.~1 through 4, and the
sources of the photometry in col.~5.

Spectral types were determined by H99 and H00 for the five reference stars,
based on spectroscopic observations obtained by their team. Because of several
issues raised in our reanalysis (see below),  we requested new observations with
the queue-scheduled Hobby-Eberly Telescope (HET; Shetrone et al.\ 2010) and its
Low-Resolution Spectrograph. These data, covering 4330--7280~\AA\ at a
resolution of 5.8~\AA, were obtained on 2013 June 12 and 13. By comparison of
our spectra with those of classification standards, we modified the spectral
types of three of the reference stars; details are given in col.~6 and a
footnote in Table~2. We then estimated the interstellar reddening, $E(B-V)$, of
each reference star by comparing the observed $B-V$ with the intrinsic $(B-V)_0$
for the indicated spectral type, as given in a literature
compilation\footnote{Available at
http://www.pas.rochester.edu/$^\sim$emamajek\slash
EEM\_dwarf\_UBVIJHK\_colors\_Teff.dat} assembled by E.~Mamajek for dwarfs, and
by Drilling \& Landolt (2000) for giants. Our adopted reddenings are given in
col.~7 of Table~2.

H99 and H00 estimated the absolute magnitude of each reference star using a
calibration of $M_V$ vs.\ spectral types. Our approach, for dwarfs, is to use
the photometric measurements, once the luminosity class and reddening have been
established from the spectral classification. We estimated the distances to
these reference stars using a calibration of the visual absolute magnitude,
$M_V$, against reddening-corrected $B-V$ and $V-I$ colors, derived through
polynomial fits to a large sample of nearby main-sequence stars with accurate
photometry and \Hipp\/ or USNO parallaxes. Our procedure is described in more
detail in Bond et al.\ (2013). This algorithm corrects for effects of
metallicity. For giants, we used the $M_V$ values tabulated by Drilling \&
Landolt (2000). Our final estimated parallaxes are listed in col.~8 of Table~2.
For comparison, the parallaxes estimated by H00 are given in col.~9.

Two of the reference stars deserve special comment. (1)~In the FGS reduction
process, the GAUSSFIT program made a substantial modification of the parallax of
REF-6, changing it from an input value near 0.7~mas to an output value of about
4~mas. This value is too large for an M0 giant at 12th mag, but if the star were
actually a misclassified M dwarf, its parallax would be another order of
magnitude even larger.  A possible explanation of the implausible FGS parallax
may be that REF-6 is a partially resolved binary with a separation slightly less
than the width of the FGS3 interference fringes ($\ltabout$40~mas). If so, the
measured position on the FGS detector can depend upon the telescope orientation,
in effect introducing a parallax-like displacement in the solution (see Nelan
2011, \S4.4.1, for examples of how close binary systems impact the observed
interference fringes). Since the FGS parallax is discrepant from any
astrophysically reasonable value, we omitted REF-6 from the solution. (2)~REF-12
was classified as a K0~V dwarf by H00. However, both its $B-V$ and $V-I$ colors
are much redder than those of a K0~V star. H00 attributed the red color to a
large interstellar reddening of $E(B-V)=0.65$. It would be surprising for such a
nearby star ($d\simeq288$~pc according to the parallax of 3.47~mas estimated by
H00) to be so reddened, given that the more distant REF-3 and REF-14 are only
moderately reddened. SS~Cyg itself is little reddened, with $E(B-V)=0.04$
according to Godon et al.\ (2012). Moreover, the {\it total\/} reddening of very
distant objects in this direction is only $E(B-V)=0.45$, according to Schlafly
\& Finkbeiner (2011). Interstellar extinction in the direction of SS~Cyg has
also been discussed by Voikhanskaya (2012), who likewise concludes that the
reddening within a few hundred parsecs is minimal. These issues would be
alleviated if REF-12 were instead a giant. Then it would be at a distance of
$\sim$5~kpc, with a parallax of $\sim$0.2~mas, and its reddening would be
$E(B-V)\simeq0.3$; such a value would be reasonably consistent with the run of
extinction vs.\ distance presented by Voikhanskaya (2012). Motivated by these
considerations, we obtained a new spectrum as described above, which confirms
that the star is a giant of about spectral type K2~III. (Unlike REF-6, REF-12
lies at the periphery of the reference-star frame, and thus the GAUSSFIT
solution is unable to put an independent tight constraint on its parallax.)

The mean parallax for the four retained reference stars according to our
estimates is 1.92~mas; in spite of the individual differences, this is fairly
close to the mean parallax of 2.23~mas adopted by H00, but actually slightly
{\it smaller}. In the analysis by H99 and H00, they forced the sum of the
parallaxes of the reference stars to be zero, and then added the mean of the
estimated parallax values to the relative parallax calculated for SS~Cyg, in
order to convert it to absolute. In our analysis, as noted above, we treat each
input reference-star parallax as an observation with an error, so that the
parallax we compute for SS~Cyg is absolute.

\subsection{Input Proper Motions of the Reference Stars}

GAUSSFIT also requires input values of the PMs of the reference stars. H99 and
H00 constrained the sum of the reference-star PMs to be zero, but in the
reanalysis by H04 they state that they used USNO-B PMs as observations with
errors. For our new analysis, we adopted input ground-based PMs as given in the
recently available UCAC4 catalog (Zacharias et al.\ 2013), which are tied to an
inertial extragalactic frame.

\subsection{Lateral Color Effect}

The FGS3 optical train includes refractive elements that introduce an
astrometrically significant color-dependent shift in observed stellar positions.
This ``lateral color effect'' must be accounted for in the analysis, especially
since SS~Cyg is blue; in particular, during outbursts, it is bluer than all of
the reference stars. Moreover, \HST\/ parallax observations require a
$\sim$$180^\circ$ change in telescope roll angle at the two extremes of parallax
factor, making the lateral shifts go in opposite directions on the sky (as does
parallactic displacement). Observations providing a calibration of lateral color
effect for FGS3 were carried out in 1994, with results reported by Benedict et
al.\ (1999), and confirmed independently by us for the present work. The
corrections are defined as functions of the $B-V$ color of each star. H99 and
H00 did not apply such corrections, but H04 did.

The $B-V$ colors of the (non-variable) reference stars are available (col.~3 of
Table~2). For SS~Cyg itself, the color varies with its magnitude level, becoming
considerably bluer during outbursts. Observations of the $B-V$ color of SS~Cyg
at various magnitude levels have been provided by Grant \& Abt (1959) and Hopp
\& Witzigmann (1980). Based on the magnitudes of SS~Cyg reported by the AAVSO
light-curve generator\footnote{Available at http://www.aavso.org/lcg} at the
exact times of the \HST\/ observations, as listed in col.~2 of Table~1, we
estimated what its $B-V$ color would have been, using the data from these two
papers. We also list, in col.~3, the $V$ magnitudes derived directly from the
FGS observations, transformed from the FGS instrumental magnitudes using the
relation of Bucciarelli et al.\ (1994); these magnitudes are in good agreement
with the AAVSO values. The implied colors are listed in col.~4 of Table~1, and
were used as input values in our FGS analysis. SS~Cyg was in outburst (and thus
very blue) during the 1997 December observations, but in quiescence (and redder)
at the other four \HST\/ visits. 

\section{Parallax and Proper Motion of SS~Cygni: VLBI and HST/FGS are
Statistically Consistent}

Our resulting PM components and absolute parallax for SS~Cygni are given in
col.~2 of Table~3. For comparison, the VLBI astrometric results are given in
col.~3, and the \HST\/ results from H00 (for the PM) and from H04 (for the
parallax) in col.~4. 

The \HST\/ observations of 1997-98 were less than optimal in two regards.
(1)~Only five reference stars were observed, although more were available in the
surrounding field. Unfortunately, one of the five caused difficulties and had to
be eliminated from our solution, as described in \S3.1. Only one of the chosen
reference stars lies to the northwest of SS~Cyg, with the other four lying to
the southeast. This limits the usefulness of a test in which each reference star
is examined as if it were the program star, in order to assess the consistency
of its input estimated parallax with the output parallax found in the solution.
(In the present case, only REF-6 could be examined in this way, as the other
reference stars were on the outer edges of the field). The small number of
reference stars also limits the utility of another standard test, in which each
star is dropped in turn from the solution to identify problem objects (e.g.,
close binaries). (2)~Observations were made at only three epochs, with the first
and third being at opposite parallax factors. It is preferable to have the
initial and final observation be made at the same parallax factor, as this
allows for a better separation between parallax and PM\null. Of course, it is
also preferable to obtain more than the minimum of three epochs needed for
parallax determination.

Nevertheless, as shown in Table~3, our reanalysis of the 1997-1998 \HST\/ data
has produced statistical agreement with the VLBI parallax. We find a parallax of
$8.30\pm0.41$~mas, differing formally by about $1.2\,\sigma$ from the VLBI value
of $8.80\pm0.12$~mas.  Our error estimate for the FGS parallax includes the
effects of uncertainties in the reference-star parallaxes. The remaining
difference in parallax between our result and that of MJ13 is plausibly a
residual systematic error, due to having only three epochs of \HST\/
observations rather than the now-standard five epochs for FGS parallax programs,
along with having only four usable reference stars.  We also note that these
older observations from 1997 and 1998 used FGS3, which has since been replaced
by the more capable and better-calibrated FGS1R.



The uncertainties listed for our PMs are the internal (random) errors. For the
PM of SS~Cyg, there will be an additional systematic  offset in the zero-point
of the PMs (i.e., a bulk linear motion of the reference frame relative to an
inertial one) due to errors in the input PMs of the reference stars. For the
UCAC4 catalog, individual PMs are uncertain at about $4\,\rm mas\,yr^{-1}$.
Thus, with four reference stars, the PM zero-point error would be of order
$2\,\rm mas\,yr^{-1}$. We find that our PM for SS~Cyg in right ascension agrees
with the VLBI value within the errors; the declination PM is slightly
discordant, at nearly $3\,\sigma$, but well within the expected systematic
error.

H00 and H04 reported smaller parallaxes ($6.27\pm0.47$~mas and
$6.06\pm0.44$~mas, respectively) of SS~Cyg. We attempted to reproduce their
parallax and PM results, by inputting the reference-star parallaxes stated in
H00 and employing the methodologies they used at the respective dates, but were
unable to do so.   The PM position angle they report differs from that in
USNO-B, UCAC4, MJ13, and the present work by about $36^\circ$. In fact, the PM
of SS~Cyg measured over six decades ago on photographic material at the Yerkes
Observatory by Strand (1948) also agrees remarkably well with the modern VLBI
and \HST\/ values in size and position angle, as does the more recent
photographic determination by Dahn et al.\ (1982). 

In summary, we find reasonable statistical agreement between our reanalysis of
the \HST/FGS parallax and the precise results from VLBI interferometry. This
reinforces our confidence that FGS astrometry does not suffer from large
systematic errors. This should be especially true when at least five epochs of
observations and a well-chosen and -calibrated reference frame are used.

\acknowledgments

This study was partially supported by NASA through grant GO-10912 from the Space
Telescope Science Institute, which is operated by AURA, Inc., under NASA
contract NAS~5-26555. We thank Bruce Margon for challenging us to reaffirm
faith in FGS parallaxes, and for useful comments. Richard Wade provided a
historical perspective.  We made use of the AAVSO Photometric All-Sky Survey
(APASS), funded by the Robert Martin Ayers Sciences Fund. We benefitted from
the  online tables of stellar data assembled by Eric Mamajek. We thank Donald
Schneider for awarding Director's Discretionary time on the HET, and the HET
Resident Astronomers for obtaining observations on short notice. The HET is a
joint project of the University of Texas at Austin, Pennsylvania State
University, Stanford University, Ludwig-Maximillians-Universit\"at M\"unchen,
and Georg-August-Universit\"at G\"ottingen. HET is named in honor of benefactors
W. P. Hobby and R. E. Eberly. The Marcario Low-Resolution Spectrograph is named
for Mike Marcario, who fabricated several optics for the instrument but died
before its completion; it is a joint project of the HET partnership and the
Instituto de Astronom\'{\i}a de la Universidad Nacional Aut\'onoma de M\'exico.

{\it Facilities:} \facility{AAVSO}, \facility{HET}, \facility{HIPPARCOS},
\facility{HST (FGS)}


\begin{deluxetable}{lccc}
\setlength{\tabcolsep}{0.2in}
\tablewidth{0pt}
\tablecaption{\HST\/ Observation Log and SS~Cyg Magnitude and Color}
\tablehead{
\colhead{Date and UT$\rm^a$} & \colhead{Visual mag$\rm^b$} &
\colhead{$V$ mag$\rm^c$} & \colhead{$B-V\rm^d$} 
}
\startdata
1997 Jun 07 16:04           &  12.1 & 12.3 & 0.6 \\
\phantom{1997} Jun 15 20:57 &  12.1 & 12.1 & 0.6 \\
\phantom{1997} Dec 12 12:01 &  8.4  &  8.6 & \llap{$-$}0.05 \\
\phantom{1997} Dec 18 03:29 &  9.4  &  9.4 & 0.1 \\
1998 Dec 10 03:31           &  11.9 & 12.0 & 0.6 \\
\phantom{1997} Dec 17 03:09 &  12.0 & 12.0 & 0.6 \\
\enddata
\tablenotetext{a}{UT is given for end of the \HST/FGS observing sequence.}
\tablenotetext{b}{Visual magnitude of SS~Cyg at time of \HST\/ observation, from
AAVSO Light-Curve Generator.}
\tablenotetext{c}{$V$ magnitude measured by FGS.}
\tablenotetext{d}{Color inferred from visual magnitude; see text.}
\end{deluxetable} 

\begin{deluxetable}{lcccllccc}
\setlength{\tabcolsep}{0.05in}
\tablewidth{0pt}
\tablecaption{Astrometric Reference Stars for SS Cygni}
\tablehead{
\colhead{ID} & \colhead{$V$} &
\colhead{$B-V$} & \colhead{$V-I$} & \colhead{Sources$\rm^a$} &
\colhead{Sp.~Type$\rm^b$} & \colhead{$E(B-V)$} &
\colhead{$\pi_{\rm est}$ [mas]} & \colhead{$\pi_{\rm est}$ [mas]} \\
\colhead{  } & \colhead{   } & \colhead{   } &
\colhead{     } & \colhead{     } & \colhead{       } & \colhead{       } &
\colhead{[this paper]} & \colhead{[H00]}
}
\startdata
REF-3  & 13.391 & 0.423 & 0.527 & 1,2,4,5,6 & F0 V   & 0.13 & 0.79	& 0.75 \\
REF-6  & 12.053 & 1.860 & 2.173 & 2,4,5,6   & M0 III & 0.30 & $\dots\rm^c$ & 0.84 \\
REF-7  & 10.865 & 0.523 & 0.587 & 1,2,3,4,7 & F8 V   & 0.00 & 4.36	& 4.09 \\
REF-12 & 15.189 & 1.470 & 1.639 & 2,4       & K2 III & 0.31 & 0.18	& 3.47 \\
REF-14 & 12.941 & 0.621 & 0.731 & 2,4,5     & F9 V   & 0.07 & 2.34	& 2.00 \\
\enddata
\tablenotetext{a}{Photometry sources: 1.~AAVSO chart. 2.~APASS\null. 3.~Grant \&
Abt (1959). 4.~H00. 5.~Henden \& Honeycutt (1997). 6.~Misselt (1996). 7.~Tycho
catalog (H{\o}g et al.\ 2000).}
\tablenotetext{b}{From H00 for REF-3 and REF-7, and from our new HET spectra for
the remaining three.}
\tablenotetext{c}{Omitted from solution; see text.}
\end{deluxetable} 

\clearpage

\begin{deluxetable}{llll}
\tabletypesize{\footnotesize}
\setlength{\tabcolsep}{0.2in}
\tablewidth{0pt}
\tablecaption{Astrometric Parameters for SS Cygni}
\tablehead{
\colhead{Parameter} & \colhead{\HST/FGS} &
\colhead{VLBI} & \colhead{\HST/FGS} \\
\colhead{         } & \colhead{[This paper]} &
\colhead{[MJ13$\rm^a$]} & \colhead{[H00,H04$\rm^b$]}}
\startdata
Proper motion in R.A.~[$\rm mas\,yr^{-1}$] \dotfill         & $112.23\pm0.75$ & 
  $112.42\pm0.07$ & $73.9\pm0.4$ \\
Proper motion in Declination [$\rm mas\,yr^{-1}$] \dotfill & $35.24\pm0.65$  & 
  $33.38\pm0.07$  & $98.0\pm1.2$ \\
Absolute Parallax [mas] \dotfill         & $8.30\pm0.41$ & 
  $8.80\pm0.12$ & $6.06\pm0.44$ \\
\enddata
\tablenotetext{a}{Miller-Jones et al.\ (2013)}
\tablenotetext{b}{Proper motions from Harrison et al.\ (2000), parallax from
Harrison et al.\ (2004)}
\end{deluxetable} 

\end{document}